\documentclass[12pt]{iopart}
\usepackage{graphicx}
\begin{document}
\title{Reaction-diffusion fronts
with inhomogeneous initial conditions}
\author{I Bena$^1$, M Droz$^1$, K Martens$^1$ and Z R\'acz$^2$}
\address{$^1$ D\'epartement de Physique Th\'eorique,
Universit\'e de Gen\`eve, CH-1211 Gen\`eve 4. Switzerland}
\address{$^2$ Institute for Theoretical Physics, E\"otv\"os 
University, 1117 Budapest, Hungary} 

\begin{abstract}

Properties of reaction zones resulting from $A+B\rightarrow C$ type
reaction-diffusion processes are investigated by analytical and numerical methods. The reagents $A$ and $B$ are separated initially and, in addition,
there is an initial macroscopic inhomogeneity in the distribution of
the $B$ species. For simple two-dimensional geometries, exact analytical
results are presented for the time-evolution of the
geometric shape of the front. We also show using cellular automata
simulations that the
fluctuations can be neglected both in the shape and in the width of the
front.

\end{abstract}
\pacs{82.20.-w,02.50.-r, 82.40.Ck}
\submitto{\JPC}
\maketitle

\section{Introduction}

Front propagation has attracted  considerable
research activity during the past years. A
large number of phenomena in different fields (e.g. crystal growth, 
chemical pattern formation, and biological invasion
problems) are determined by the properties of various types of fronts.

An important class of problems is related to the case of front propagation
into unstable states~\cite{vansaarloos}.
They describe situations  in which the dynamics of a system
is governed by a front which invades a domain where the system
is in an unstable state,
leading to the onset of a new stable  state (often with spatio-temporal
structures present).
Examples include population dynamics
modeled by the extended Fisher-Kolmogorov equation~\cite{FK},
onset of vortices in
Taylor-Couette flows~\cite{RB} and of convection rolls
in Rayleigh-B\'enard experiments
(modeled by the Swift-Hohenberg equation~\cite{SH}),
phase turbulence in autocatalytic chemical
reactions modeled by the Kuramoto-Sivashinsky equation~\cite{KS}.

Most of the studies have been
addressing the properties of reaction-diffusion
fronts in homogeneous media~\cite{Cross-Hoh}.
Presently, however, an
increasing importance is attached to
the more complicated cases of fronts
propagating in {\sl inhomogeneous media}.
Inhomogeneities may have different natures:
chemical (related to spatial inhomogeneities in the concentration of the reagents),
physical (reagents' diffusion coefficients may vary in space),
geometrical (changes
in the geometry of the medium), etc. Also,
inhomogeneities may be either related to an intrinsic disorder
present in the system, or may be created artificially.
There have been both
theoretical and experimental effort
in studying various situations, see e.g.~\cite{xin,mendez,zha,markus,sainhas}.
Examples are numerous
in the context of biology~\cite{bio}, population dynamics~\cite{popdyn},
heterogeneous catalysis~\cite{cata}, or heterogeneous excitable media like
the photosensitive Belousov-Zhabotinsky reaction~\cite{belousov,oosawa,petrov}.
In particular, it was argued that under some circumstances
reaction-diffusion fronts may behave like light-waves
when undergoing refraction and reflection~\cite{zha,markus,sainhas}.

In the present paper we shall be concerned with a different class of
phenomena which emerges when
the unstable state is produced by reaction processes in a front moving
diffusively. In the wake of the front, the dynamics of the unstable
state may generate various spatio-temporal patterns. A prototype
of this pattern-forming process is the Liesegang phenomena~\cite{henisch}\footnote{
While the reagents in Liesegang phenomena are chemicals, examples where
the reagents are of more exotic nature (such as topological defects \cite{kotomin}) are also known.} where
a chemical reactant
(called inner electrolyte) is dissolved in a gel matrix  and a second reactant
(outer electrolyte) diffuses in and reacts with the inner electrolyte.
Under certain conditions, the dynamics of the reaction product generates
a family of high-density precipitate zones whose properties
obey generic laws~\cite{mataus} which have been understood in terms of
a simple phase separation scenario \cite{scenario}.

Most of the Liesegang pattern studies have
been made in homogeneous conditions~\cite{henisch},
when the initial concentration of the inner electrolyte is uniform.
In recent work by Grzybowski {\em et al}~\cite{grzyb}, however, the
experiments were performed in inhomogeneous conditions (with a step-like
profile in the height of the gel containing the inner electrolyte).
It was argued that Liesegang bands emerging in the two regions of different
thicknesses join each other at the step according to
a Snell law, where the indices of refraction are related
to the so-called spacing coefficients of the patterns.
 
In order to develop a theoretical basis for the above findings,
we shall first investigate the properties
of the diffusive reaction front in the presence of inhomogeneities
in the initial distribution of the inner electrolyte. This is the simplest
way to take into account the inhomogeneities and it allows to
obtain detailed understanding and analytical solutions.
The ultimate goal, left for further studies, will be the use of
the reaction zone results as an input to the description of the
band formation in the
spinodal decomposition scenario~\cite{scenario} which has been shown
to correctly describe
the pattern formation in the homogeneous case.

The paper in organized as follows. In Sec.~2, we present a brief
review of the main results known about the properties of diffusive
reaction fronts in case when the electrolytes are initially separated
but otherwise are homogeneously distributed.
Section~3 is devoted to the study of fronts when inhomogeneity is
present in the initial state of the inner electrolyte.
Both analytical results, at the mean-field level,
as well as numerical results are given.
Cellular automata simulations are presented in
Sec.~4 and the role of fluctuations is discussed.

\section{Reaction front: Homogeneous distribution of the inner electrolyte}

We review here the properties of reaction front emerging
in an $A+B \to C$
process where the particles $A$ and $B$ diffuse and react with rate $k$ upon
contact. Initially, the reagents are separated, with the inner electrolyte
$B$ occupying the $x>0$ half-space (homogeneously distributed with concentration $b_0$), while the outer
electrolyte $A$ is homogeneously
distributed with concentration $a_0$ in the $x<0$ half-space.
Usually, the case $a_0\gg b_0$ is considered and one finds
that $A$ and $B$ react in a confined region (the reaction front) which
moves in the positive $x$ direction. The properties of this front can be addressed at different levels:
a mean-field like one, neglecting fluctuations, or approaches that take
the fluctuations into account. 

\subsection{Mean-field analysis}

The mean-field study of the front resulting from the reaction $A+B \to C$ has
been made by G\'alfi and R\'acz~\cite{GR}. The generalization to the reaction 
$nA+mB \to C$ has been given by Cornell {\em et al}~\cite{corchop}.
We shall recall here the main results without going into the details of
the calculations.

In the homogeneous initial condition case and with the fluctuations neglected,
the concentration of the reagents $a(x,t)$ and $b(x,t)$ will depend only  on the variable $x$ at all times
and the problem becomes one-dimensional. The reaction-diffusion equations
for the concentrations $a(x,t)$ and $b(x,t)$ read:
\begin{equation}
\partial_ta =D_a \partial_x^2a -kab\,, \quad \quad \quad \partial_t b =
D_b \partial_x^2b -kab\,,
\label{rd-eq}
\end{equation}
where $k$ is the reaction rate and $D_a, D_b$ are, respectively, the diffusion
coefficients of $A$ and $B$.

The production rate of $C$, which is a quantity of interest since it is the
source for the precipitation process in Liesegang phenomena \cite{scenario},
is obtained by computing $R(x,t)=ka(x,t)b(x,t)$. Actually, the region where
$R(x,t)$ is significantly different from zero defines the reaction zone.
The center of the zone, $x_f(t)$, can be found from the condition
$a(x_f,t)=b(x_f,t)$, and for the case of $D_a=D_b=D$ this leads to
especially simple results. Indeed, one notices then that
$u(x,t)=a(x,t)-b(x,t)$ obeys a simple diffusion equation. Introducing
$q=b_0/a_0$, the solution
is  given in  terms of the error function ${\rm erf}$,
\begin{equation}
u(x,t)=a_0 \left[\frac{1-q}{2}-\frac{1+q}{2} \,{\rm erf}\left(\frac{x}{2\sqrt{D t}}\right)\right]
\label{eq1}
\end{equation}
and the condition $a(x_f,t)=b(x_f,t)$ yields
\begin{equation}
x_f(t)=\sqrt{2D_ft}\;.
\label{x_f}
\end{equation}
This means that the front moves diffusively and its
diffusion coefficient $D_f$ is  given by the equation
${\rm erf}(\sqrt{D_f/2D})={(1-q)}/{(1+q)}$.

In order to complete the description of the front one needs the solution
of the nonlinear part of the reaction-diffusion equations,
\begin{equation}
\partial_t a=D \partial_x^2 a -k(a^2+au)\,.
\end{equation}
It is not possible to obtain a general solution of this equation. However, assuming 
that the width $w(t)$ of the reaction front, defined as
\begin{equation}
w^2(t)=\frac{\displaystyle\int_{-\infty}^\infty (x-x_f)^2 R(x,t) dx}
{\displaystyle\int_{-\infty}^\infty  R(x,t) dx}\;,
\end{equation}
is much smaller that the diffusion length $\ell_{\mbox{diff}}(t)=\sqrt{2Dt}$
(an assumption that is justified a posteriori),
one finds that, in the
long-time limit,
the quantities of interest take the following scaling forms
in the reaction zone region:
\begin{eqnarray}
&& w(t) \sim t^\alpha\,,\quad a(x,t)=t^{-\gamma} {\hat a}(xt^{-\alpha})\,,\nonumber\\
&& b(x,t)=t^{-\gamma} {\hat b}(xt^{-\alpha})\,, \quad R(x,t)= t^{-\beta}
{\hat R}(xt^{-\alpha})\, .
\label{scalingforms}
\end{eqnarray}
Here the exponents are given as $\alpha=1/6$, $\beta=2/3$ and $\gamma=1/3$.
The exponents actually obey two general scaling
relations following from the facts that $\ell_{\mbox{diff}}(t)\gg w(t)$ and
that the reaction zone is fed by the fluxes of diffusing particles \cite{GR,corchop},
\begin{equation}
\alpha +\gamma =1/2\,,  \quad \quad \quad 2\alpha+\gamma=\beta\,.
\end{equation}
In the mean-field approximation, one has an extra scaling relation, namely $\beta=2\gamma$, and this leads to the above values of the exponents.
Note that the case $D_a \not= D_b$ is more complicated. However,
the exponents obey
the same scaling relations, and only the scaling functions ${\hat a}, {\hat b}$, 
and ${\hat R}$ are affected.

A similar analysis can be carried out for the more general reaction 
$nA+mB \to C$~\cite{corchop}. The main modification arises when expressing the production  
rate,  and one has now the scaling relation $\beta=(n+m)\gamma$, 
leading to the mean-field  exponents:
\begin{equation}
\alpha=\frac{n+m-1}{2(n+m+1)}, \quad \beta=\frac{n+m}{n+m+1}, \quad \gamma=\frac{1}{n+m+1}\,.
\end{equation}

\subsection{Beyond mean-field}

Numerical simulations have been performed for the general reaction
$nA+mB \to C$~\cite{corchop},
for several values of $(n,m)$, including the case $n=m=1$, and
confirm the validity
of the above scaling relations. Moreover, the mean-field exponents
are recovered in dimensions larger than $d=2$.
In $d=1$, the exponent values strongly differ from their mean-field
values. This leads to the hypothesis that the upper critical
dimension for diffusive reaction fronts is $d_u=2$.

The validity of this hypothesis has been supported by theoretical arguments (Cornell and 
Droz~\cite{cordro}) based on the scaling  behavior of the steady-state front generated by imposing antiparallel fluxes of $A$ and $B$ particles.
This situation is much easier to investigate, since the front is time-independent,
and depends only on three relevant parameters, the flux $J$, the diffusion coefficient
$D$, and the reaction rate $k$. The results, however, can be directly
applied to the time-dependent case too, since the front is
formed quasistatically by diminishing fluxes
$J(t) \sim t^{-1/2}$. Assuming that scaling applies even when
fluctuations are present
(as seen in numerical simulations), a dimensional analysis and
scaling arguments lead to the following conclusions: first, there is an upper critical 
dimension $d_u=2/(m+n-1)$ above which mean-field predictions are correct; second, for $d <d_u$
(only realizable for $m=n=1$), the width exponent is $\alpha=1/2(d+1)$, which for $d=1$
gives $\alpha=1/4$ instead of $\alpha=1/6$ for the mean-field case.

In summary, one finds that, in the physically relevant dimensions,
the reaction zone moves diffusively and its
width is negligible with respect to the diffusion length of the problem.
There is a third result which has not been discussed here
but will be important
in a later application when the front will figure as a source of particles
in the description
of the formation of Liesegang bands. Namely, the front leaves behind
a constant concentration of $C$ particles \cite{GR,mataus}.

\section{Reaction front: Inhomogeneous distribution of the inner electrolyte}

\begin{figure}[h!]
\begin{center}
\quad\vspace{1.5cm}\\
\includegraphics[angle=0, width=0.5\columnwidth]{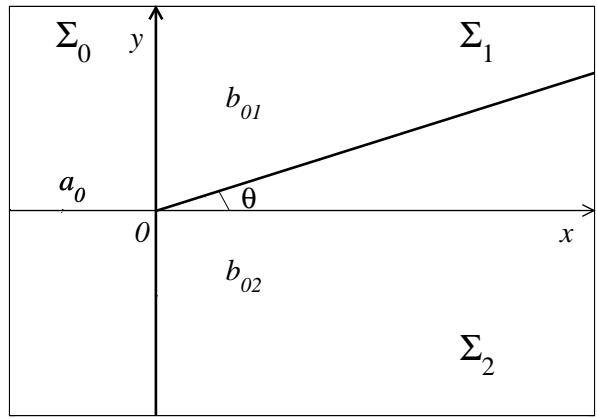}
\end{center}
\caption{\small{Geometry of the inhomogeneous media considered.}}
\label{geo}
\end{figure}

There are many ways to introduce inhomogeneities. We shall consider
here a simple case which, on one hand, can be treated analytically and,
on the other hand, mimics
a system with a geometrical inhomogeneity quite similar
to the experimental setup used
in ~\cite{grzyb}. Namely, we shall assume that the inhomogeneity is in
the initial concentration of the inner electrolyte. As described in
Fig.\ref{geo}, the reagents are separated in the
plane $(x,y)$ into three zones, $\Sigma_0, \,\Sigma_1$, and $\Sigma_2$.
For $x<0$ and $\forall y$, $a=a_0;~ b=0$ (region $\Sigma_0$),
while for $x>0$ the
half-plane is divided in two domains, separated by a line making an angle $\theta$
with the $x$ axis. In the upper domain $\Sigma_1$, $a=0;~ b=b_{01}$,
while in the lower domain $\Sigma_2$, $a=0;~ b=b_{02}$.
This corresponds to an engineered initial
inhomogeneity of the reagent $B$.

As before we restrict ourselves to the reaction $A+B \to C$,
although the results could be extended to the more general case
$mA+nB \to C$. We also assume that the diffusion coefficients are equal $D_a=D_b=D$.

\subsection{Mean-field description of the reaction front}

The mean-field description is again in terms of the reaction-diffusion
equations (\ref{rd-eq}), the only difference being that the concentrations
depend on two spatial variables ($x$, $y$) and the diffusion terms
are now expressed through the two-dimensional Laplacian.

The position of the reaction front $(x_f(t)\,,y_f(t))$ is again obtained
by solving the diffusion equation for $u(x,y,t)=a(x,y,t)-b(x,y,t)$
and finding $(x_f,y_f)$ satisfying the condition $u(x_f,y_f,t)=0$.
The diffusion equation being linear, the solution
$u(x,t)$ for the initial condition in Fig.\ref{geo}
can be constructed as a superposition of two particular
solutions $u=u_1+u_2$.  Here
$u_1(x,y,t)$ satisfies the initial condition $u_1(x<0,y,t=0)=a_0$
and $u_1(x>0,y,t=0)=-b_{01}$ and, according
to Eq.~(\ref{eq1}), is given by
\begin{equation}
u_1(x,y,t)=a_0 \left[\frac{1-q_1}{2}-\frac{1+q_1}{2}\, 
{\rm erf}\left(\frac{x}{2\sqrt{D t}}\right) \right]\quad
\end{equation}
where $q_1=b_{01}/a_0$.

The second function $u_2(x,y,t)$ satisfies the initial condition
$u_2=0$ in $\Sigma_0$ and $\Sigma_1$,
and $u_2=b_{01}-b_{02}=a_0(q_1-q_2)$ in $\Sigma_2$ with $q_2=b_{02}/a_0$.
The solution is constructed in terms of the Green function for the diffusion operator
\begin{equation}
G(x,y,t)=\frac{1}{4\pi Dt}\exp\left( -\frac{x^2 +y^2}{4Dt} \right)
\end{equation}
as an integral restricted to the region $\Sigma_2$
\begin{equation}
u_2(x,y,t)= a_0(q_1-q_2) \int_{\Sigma_2} dx' dy' G (x-x', y-y', t)\,.
\end{equation}

\begin{figure}
\begin{center}
\quad \vspace{1.5cm}\\
\includegraphics[angle=0, width=0.85\columnwidth]{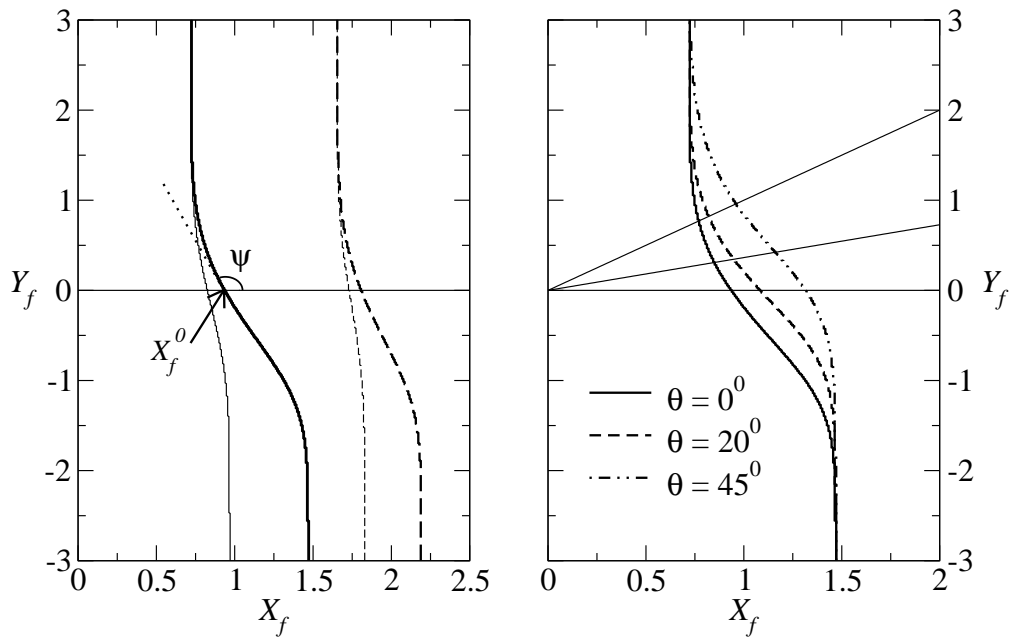}
\end{center}
\caption{\small{Shape of the reaction-diffusion front (in scaled time-independent variables)
for different values of the parameters. Left panel: front shape for $\theta=0$. 
The thick continuous line corresponds to $q_1=0.2$ and $q_2=0.02$;
the thin continuous line to $q_1=0.2$ and $q_2=0.1$; thick dashed line: $q_1=0.01$, $q_2=0.001$;
and thin dashed-line: $q_1=0.01$, $q_2=0.005$.
$X_f^0$ gives the intersection with the $x$ axis, and $\psi$ is the angle between the tangent at $X_f^0$ and the $x$ axis.
Right panel: front shape
for different values of $\theta$ ($q_1=0.2$ and $q_2=0.02$).
The thin straight lines indicate, respectively, the
border between the regions $\Sigma_1$ and $\Sigma_2$  for 
the considered values of $\theta$.}}
\label{res_io}
\end{figure}

For simple cases, the integration can be performed analytically, leading to
an explicit expression for $u_2$.
For example, for the symmetric case when $\theta=0$, one obtains
\begin{eqnarray}
u(x,y,t)&=&a_0\left\{\frac{1-q_1}{2} -\frac{1+q_1}{2}\,{\rm erf}\left(\frac{x}{2\sqrt{Dt}}\right)\right.\nonumber\\
&&\left.+\frac{q_1-q_2}{4} \left[1 + {\rm erf}\left(\frac{x}{2\sqrt{Dt}}\right)\right]
\,\left[1-{\rm erf}\left(\frac{y}{2\sqrt{Dt}}\right)\right]\right\}\,.
\end{eqnarray}
The position of the reaction front is thus given by the implicit equation:
\begin{equation}
\frac{1-q_1}{2} -\frac{1+q_1}{2}\,
{\rm erf}(X_f)+\frac{q_1-q_2}{4} \left[1 + {\rm erf}(X_f)\right]
\,\left[1-{\rm erf}(Y_f)\right]=0\,,
\end{equation}
where
\begin{equation}
X_f=\frac{x_f(t)}{2\sqrt{Dt}}\,,\quad Y_f=\frac{y_f(t)}{2\sqrt{Dt}}\,.
\end{equation}
are the scaled, {\em time-independent} coordinates of the front.
The corresponding shape of the front is shown
in the left panel of Fig.\ref{res_io}
for different values of the parameters $q_1$ and $q_2$.

Far from the $x$ axis ($|y|\gg \sqrt{2Dt}$), the reaction
front is not affected by the inhomogeneity, it is parallel to the $y$ axis
and moves diffusively,
along the $x$ direction. The diffusion coefficients are given,
respectively, by $\mbox{erf}(\sqrt{D_f^{1,2}/2D})=(1-q_{1,2})/(1+q_{1,2})$.

Near the $x$ axis, there is a crossover region between the two almost
linear segments of the front. The crossover is smooth and can be
characterized by the
front position $x_f^0(t)=2X_f^0\sqrt{Dt}$ on the $x$ axis and by the
angle $\psi$ between the tangent to the front at $X_f^0$ and the
$x$ axis. The front position on the $x$ axis is obtained from
\begin{equation}
{\rm erf}(X_f^0)= \frac{1-({q_1+q_2})/{2}}{1+({q_1+q_2})/{2}}\,,
\end{equation}
and, of course, it also moves diffusively.
The angle $\psi$ on the other hand is a constant
\begin{equation}
\tan \psi = -\frac{[1+(q_1+q_2)/2]^2}{q_1-q_2}\exp[-(X_f^0)^2]\,.
\end{equation}

A similar calculation can be performed for an arbritrary angle $\theta$, however 
the results can no longer be expressed in terms of explicit functions. Namely, 
the position of the front
(in terms of scaled variables) is given implicitly by the following equation:
\begin{eqnarray}
&&\frac{2-q_1-q_2}{4}-\frac{2+q_1+q_2}{4}\,\mbox{erf}(X_f)\nonumber\\
&&-\frac{q_1-q_2}{2\sqrt{\pi}}\int_{-\infty}^{X_f} dz \exp(-z^2)\,\mbox{erf}
(Y_f-X_f\,\mbox{tan}\theta - z\,\mbox{tan}\theta)=0\,.
\end{eqnarray}
The profile of the front is given in the right panel of Fig.~2 for different
values of the angle $\theta$. The analytical results have been checked by numerical integration
of the reaction-diffusion equations for the reagents $A$ and $B$.
Note that the infinite reaction
rate $k$ limit is especially convenient for carrying out the
numerical work since the position of the front does
not depend on $k$ for the $D_a=D_b$ case. However, it does not allow the
investigation of the width of the front. For finite $k$,
the width of the front in the asymptotic regions $Y_f\to \pm\infty$
follows from the case of homogeneous initial distribution of inner electrolyte,
i.e. $w(t)\sim t^\alpha$ with $\alpha=1/6$. Investigations of
the crossover region in the case when fluctuations are
also included (see next section)
indicate that no anomalous behavior occurs.
Thus we expect that the width
remains much smaller than the diffusion length and the scaling
$w(t)\sim t^\alpha$ with $\alpha=1/6$ remains valid everywhere
along the front.

The conclusions one can draw from the analytic results and from
Fig.\ref{res_io} are as follows. First, the front motion remains diffusive
but the shape deviates from straight line. Second, the distortion is smooth
and can be parameterized by the two asymptotic limits at $Y_f\to \pm\infty$,
and by a linear parameterization of the crossover region.
\subsection{Beyond mean-field }

To test whether the mean-field predictions are robust against including
fluctuations, we will compare the above results to those
given by a mesoscopic approach, namely by the simulation of the reaction-diffusion front by a cellular automata. This approach
also provides informations about how the width of the front scales with time,
since a finite reaction rate is considered.

The model (described  in more detail
in \cite{Cho94}) is defined on a two-dimensional square lattice
of $250 \times 250$ sites with synchronous dynamics
at discrete time steps. One iteration of the automata is defined as the combination of one diffusion and one reaction step.
The diffusion coefficients for A and B particles
are chosen to be equal, and the reaction $A + B \to C$
takes place only for particles entering a site from opposite directions.
We use thus
a finite reaction rate, since the probability for a reaction of two particles
is equal to $1/4$. The front position is defined as the region where the difference
of the two concentrations, computed in a Moore neighborhood
for the A and B particles, vanishes. To allow for better
statistics the results are averaged over 750 runs.

The initial state is prepared as illustrated in Fig.\ref{geo}.
In the region $\Sigma_0$ (with a length of 100 sites), all the
sites are fully occupied, i.e. the concentration of $A$'s is
$a_0=1$. The regions $\Sigma_1$ and $\Sigma_2$
(of length of 150 sites)
are randomly occupied by B particles
with concentration $b_{01}=0.2$ and $b_{02}=0.02$, respectively.

\begin{figure}
\begin{center}
\quad\vspace{1.5cm}\\
\includegraphics[width=0.9\columnwidth]{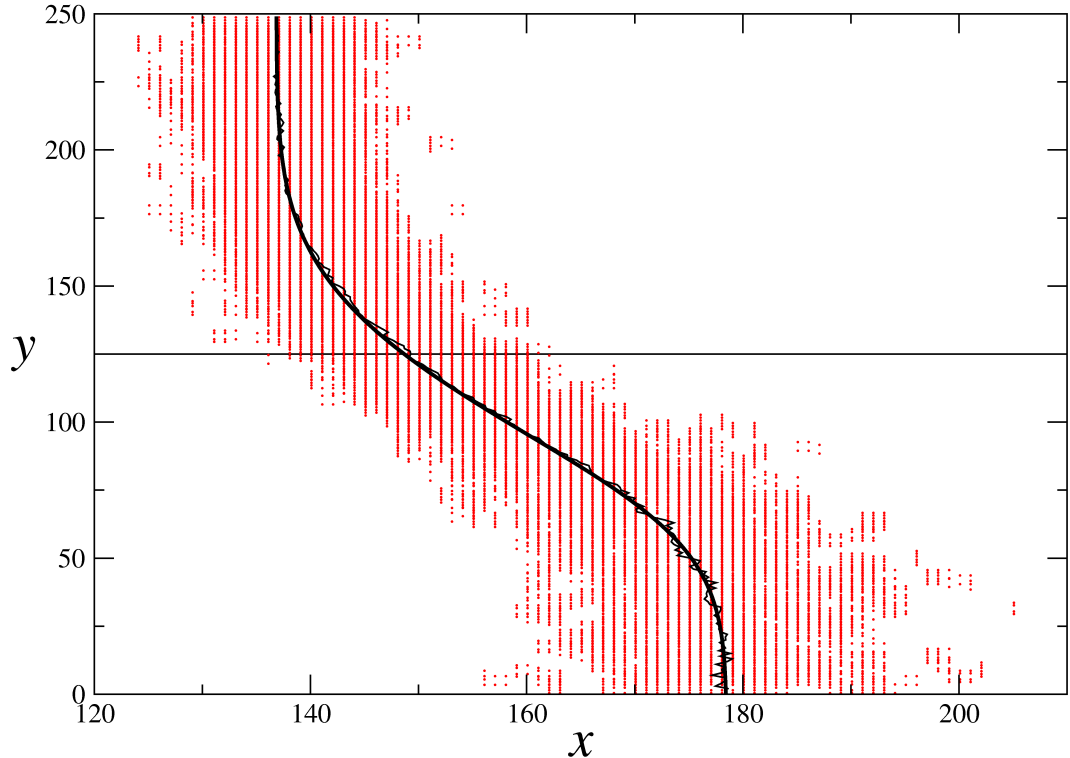}
\end{center}
\caption{\small{Results of the cellular automata simulations
of the reaction-diffusion front for $\theta=0$. The values of the initial
concentrations are $a_0=1$, $b_{01}=0.2$ and $b_{01}=0.02$, and the diffusion
coefficients for the A and B particles are equal $D_a=D_b$.
The straight horizontal line separates the high- and the
low-concentration regions of B particles. The points in the dotted
region represent the instantaneous position of the front,
calculated from condition $u=a-b=0$ after 3000 timesteps.
The thin noisy line gives the
position of the front averaged over 750 realizations of the
process. The analytic result is shown by the thick line.}}
\label{res_ki}
\end{figure}

The results of the cellular automata
simulations for the case of $\theta=0$ are shown in Fig.\ref{res_ki}.
We find good agreement between the analytical results and the averaged
position of the front obtained from the cellular automata approach. The
fluctuations in the system therefore appear to play no role for
the position of
the front. Equally good agreement was found for the case with
a nonzero angle $\theta=20^0$.

Preliminary investigations of the time dependence of the
width of the front at the separation
line between $\Sigma_1$ and $\Sigma_2$ seems to indicate 
the same time scaling $w(t)\sim t^{1/6}$
as in the homogeneous case (as mentioned before,
far from the inhomogeneity area
the behavior is obviously the one following from homogeneous initial
distribution of the inner electrolyte).

\section{Discussion}

We have considered the motion of the reaction zone for a case of simple
initial-state inhomogeneity
which mimics the experimental setup where the Snell law had been
discussed for Liesegang bands \cite{grzyb}. Our main finding is that although the
straight-line shape of the front becomes distorted due to the
inhomogeneity, the basic properties such as (1) the diffusive overall
motion, (2) the negligible width, and (3) the irrelevance of
fluctuations in the physical dimensions remain unchanged.

The next step towards understanding whether a version of the Snell
law is valid for Liesegang bands meeting at a line of inhomogeneity
should be the use of our present results as an input for a theory
of Liesegang band formation. The best candidate should be the
spinodal decomposition description which has been shown \cite{scenario}
to reproduce all the known generic properties of the Liesegang
patterns.

\section*{Acknowledgement}

This research has been partly supported by the
Swiss National Science Foundation and by the Hungarian Academy
of Sciences (Grants No.\ OTKA T043734 and TS 044839).

\section{Bibliography}


\end{document}